\PassOptionsToPackage{unicode}{hyperref}
\PassOptionsToPackage{hyphens}{url}
\PassOptionsToPackage{dvipsnames,svgnames,x11names}{xcolor}
\documentclass[
]{article}

\usepackage{amsmath,amssymb}
\usepackage{iftex}
\ifPDFTeX
  \usepackage[T1]{fontenc}
  \usepackage[utf8]{inputenc}
  \usepackage{textcomp} 
\else 
  \usepackage{unicode-math}
  \defaultfontfeatures{Scale=MatchLowercase}
  \defaultfontfeatures[\rmfamily]{Ligatures=TeX,Scale=1}
\fi
\usepackage{lmodern}
\ifPDFTeX\else
\fi
\IfFileExists{upquote.sty}{\usepackage{upquote}}{}
\IfFileExists{microtype.sty}{
  \usepackage[]{microtype}
  \UseMicrotypeSet[protrusion]{basicmath} 
}{}
\makeatletter
\@ifundefined{KOMAClassName}{
  \IfFileExists{parskip.sty}{%
    \usepackage{parskip}
  }{
    \setlength{\parindent}{0pt}
    \setlength{\parskip}{6pt plus 2pt minus 1pt}}
}{
  \KOMAoptions{parskip=half}}
\makeatother
\usepackage{xcolor}
\ifx\paragraph\undefined\else
  \let\oldparagraph\paragraph
  \renewcommand{\paragraph}[1]{\oldparagraph{#1}\mbox{}}
\fi
\ifx\subparagraph\undefined\else
  \let\oldsubparagraph\subparagraph
  \renewcommand{\subparagraph}[1]{\oldsubparagraph{#1}\mbox{}}
\fi

\usepackage{longtable,booktabs,array}
\usepackage{calc} 
\usepackage{etoolbox}
\makeatletter
\patchcmd\longtable{\par}{\if@noskipsec\mbox{}\fi\par}{}{}
\makeatother
\IfFileExists{footnotehyper.sty}{\usepackage{footnotehyper}}{\usepackage{footnote}}
\makesavenoteenv{longtable}
\usepackage{graphicx}
\makeatletter
\def\maxwidth{\ifdim\Gin@nat@width>\linewidth\linewidth\else\Gin@nat@width\fi}
\def\maxheight{\ifdim\Gin@nat@height>\textheight\textheight\else\Gin@nat@height\fi}
\makeatother
\setkeys{Gin}{width=\maxwidth,height=\maxheight,keepaspectratio}
\makeatletter
\def\fps@figure{htbp}
\makeatother
\NewDocumentCommand\citeproctext{}{}

\makeatletter
 \let\@cite@ofmt\@firstofone
 \def\@biblabel#1{}
 \def\@cite#1#2{{#1\if@tempswa , #2\fi}}
\makeatother
\newlength{\cslhangindent}
\newlength{\csllabelwidth}
\newenvironment{CSLReferences}[2] 
 {\begin{list}{}{%
  \setlength{\itemindent}{0pt}
  \setlength{\leftmargin}{0pt}
  \setlength{\parsep}{0pt}
  \ifodd #1
   \setlength{\leftmargin}{\cslhangindent}
   \setlength{\itemindent}{-1\cslhangindent}
  \fi
  \setlength{\itemsep}{#2\baselineskip}}}
 {\end{list}}
\usepackage{calc}

\usepackage{arxiv}
\usepackage{orcidlink}
\usepackage{amsmath}
\usepackage[T1]{fontenc}
\makeatletter
\@ifpackageloaded{caption}{}{\usepackage{caption}}
\AtBeginDocument{%
\ifdefined\contentsname
  \renewcommand*\contentsname{Table of contents}
\else
  \newcommand\contentsname{Table of contents}
\fi
\ifdefined\listfigurename
  \renewcommand*\listfigurename{List of Figures}
\else
  \newcommand\listfigurename{List of Figures}
\fi
\ifdefined\listtablename
  \renewcommand*\listtablename{List of Tables}
\else
  \newcommand\listtablename{List of Tables}
\fi
\ifdefined\figurename
  \renewcommand*\figurename{Figure}
\else
  \newcommand\figurename{Figure}
\fi
\ifdefined\tablename
  \renewcommand*\tablename{Table}
\else
  \newcommand\tablename{Table}
\fi
}
\@ifpackageloaded{float}{}{\usepackage{float}}
\floatstyle{ruled}
\@ifundefined{c@chapter}{\newfloat{codelisting}{h}{lop}}{\newfloat{codelisting}{h}{lop}[chapter]}
\floatname{codelisting}{Listing}

\makeatother
\makeatletter
\@ifpackageloaded{caption}{}{\usepackage{caption}}
\@ifpackageloaded{subcaption}{}{\usepackage{subcaption}}
\makeatother
\ifLuaTeX
  \usepackage{selnolig}  
\fi
\usepackage{bookmark}

\IfFileExists{xurl.sty}{\usepackage{xurl}}{} 
\hypersetup{
  pdftitle={This is not normal! (Re-) Evaluating the lower n guidelines for regression analysis},
  pdfauthor={David Randahl\^{}*},
  pdfkeywords={Number of observations, Normality, Linear
regression, Central limit theorem, Simulation},
  colorlinks=true,
  linkcolor={blue},
  filecolor={Maroon},
  citecolor={Blue},
  urlcolor={Blue},
  pdfcreator={LaTeX via pandoc}}

\newcommand{\runninghead}{A Preprint }
\renewcommand{\runninghead}{A Preprint }
\title{This is not normal! (Re-) Evaluating the lower n guidelines for
regression analysis\thanks{The research was funded by the European
Research Council, project H2020-ERC-2015-AdG 694640 (ViEWS) and
Riksbankens Jubileumsfond, grant M21-0002 (Societies at Risk)}}
\author{\textbf{David
Randahl\(^*\)}~\orcidlink{0000-0003-1069-6067}\\Department of Peace and
Conflict Research\\Uppsala
University\\\\\href{mailto:david.randahl@pcr.uu.se}{david.randahl@pcr.uu.se}}
\date{}
\begin{document}
\maketitle
\begin{abstract}
The commonly cited rule of thumb for regression analysis, which suggests
that a sample size of \(n \geq 30\) is sufficient to ensure valid
inferences, is frequently referenced but rarely scrutinized. This
research note evaluates the lower bound for the number of observations
required for regression analysis by exploring how different
distributional characteristics, such as skewness and kurtosis, influence
the convergence of t-values to the t-distribution in linear regression
models. Through an extensive simulation study involving over 22 billion
regression models, this paper examines a range of symmetric,
platykurtic, and skewed distributions, testing sample sizes from 4 to
10,000. The results show that it is sufficient that either the dependent
or independent variable follow a symmetric distribution for the t-values
to converge at much smaller sample sizes than \(n=30\), unless the other
variable is extremely skewed. This is contrary to previous guidance
which suggests that the error term needs to be normally distributed for
this convergence to happen at low \(n\). However, when both variables
are highly skewed, much larger sample sizes are required. These findings
suggest the \(n \geq 30\) rule is overly conservative in some cases and
insufficient in others, offering revised guidelines for determining
minimum sample sizes.
\end{abstract}
{\bfseries \emph Keywords}
\def\sep{\textbullet\ }
Number of observations \sep Normality \sep Linear
regression \sep Central limit theorem \sep
Simulation

\newpage{}

\section{Introduction}\label{introduction}

When I was a student and attended my first quantitative methodology
course at university, I was told that for regression modelling to be
appropriate we would either have to assume that the distribution of the
error term (\(\epsilon\)) was normally distributed, \emph{or} that the
sample size was large enough so that the central limit theorem would
ensure that the resulting statistics of interest would follow an
approximate normal distribution. When the professor was questioned about
the required sample size they replied that a common rule of thumb was
\(n=30\). At the time, I did not think much more about that but simply
accepted the \(n\geq30\) rule of thumb.

A decade on, however, as I was preparing to teach the same material to a
new cohort of students, I decided to demonstrate the Central Limit
Theorem in action by progressively increasing the sample size (\(n\)) of
dice throws and calculating the mean. During this exercise, I realized
that the distribution of the mean of dice throws converged to the normal
distribution at much lower \(n\) than 30. I also found that other, more
skewed, distributions took much longer to converge. Around the same
time, a group of master's students I was supervising began pestering me
for a reference on the \(n\geq30\) rule of thumb to use in their thesis.
This wasn't the first time I had been asked for this, but these students
were unusually persistent.

Certain that there would be plenty of systematic tests of this rule of
thumb, I decided to look into where the rule of thumb came from, and
what alternative rules of thumb existed. To my surprise, while criticism
of the \(n\geq30\) rule of thumb was rife, few systematic tests of this
(or other) rules of thumb for regression analysis seemed to exist. In
addition, with one notable exception, the tests that did exist did not
focus on the distributional assumptions of the regression analysis but
rather on other (important) characteristics such as power. In this
research note I aim to critically examine the lower bound for the number
of observations in regression analysis using modern simulation methods,
and to provide guidance on the lower bound of \(n\) for linear
regression analyses. The results show that we should likely revise both
what we teach about the lower bound of observations in regression as
well as under what circumstances regression analysis is appropriate.

\section{The distributional assumption of linear
regression}\label{the-distributional-assumption-of-linear-regression}

The classical linear regression model assumes that the dependent
variable \(Y\) is a linear function of the independent variable(s) \(X\)
and an error term \(\epsilon\): \(Y = \beta_0 + \beta_1X + \epsilon\).
The parameters are commonly estimated using the ordinary least squares
(OLS) method, which minimizes the sum of the squared residuals. Under
the condition that the error term, \(\epsilon\), is normally
distributed, the resulting estimated coefficients,
\(\hat{\pmb{\beta}}\), follow a normal distribution in repeated
sampling. When estimating the standard errors of the estimated
coefficients and standardizing them, the resulting quantity,
\((\hat{\beta}_j - \beta_j)/\text{SE}(\hat{\beta}_j)\), follows a
\emph{t} distribution with \(n-k\) degrees of freedom in repeated
sampling, where \(k\) is the number of parameters in the model.
Inferences about the population parameters can then be made using the
\emph{t} distribution and its associated critical values or p-values.
Crucially, the inferences about the parameters are generally made by
testing the \textbf{null hypothesis} that the population parameter is
equal to zero versus the \textbf{alternative hypothesis} that the
population parameter is different from zero.\footnote{Other hypothesis
  tests are possible, but the null hypothesis that the population
  parameter is equal to zero is the most common in regression analysis.}
These tests assume that the \textbf{null hypothesis} is true, and thus
that the estimated \(t\)-values follow a \(t\)-distribution with a mean
of zero and a standard deviation of one.

In this setting, the distributional assumption of the error term
\(\epsilon\) is essential for the validity of the OLS estimator, as
without it there is no guarantee that the estimated \(t\)-values will
follow a \(t\)-distribution. If the estimated \(t\)-values do not follow
this distribution, the critical values and p-values may not be valid,
leading to incorrect inferences.

The normality assumption of linear regression can, however, be relaxed
when the sample size is sufficiently large. This is because of the
central limit theorem which ensures that when the sample size is large
\(\hat{\pmb{\beta}}\) will follow an approximate normal distribution,
and the estimated \(t\)-values will simularly follow an approximate
\(t\)-distribution, in repeated sampling, even if the error term
\(\epsilon\) is not normally distributed (Wooldridge 2012).

\section{\texorpdfstring{Guidelines for \emph{n} in linear
regression}{Guidelines for n in linear regression}}\label{guidelines-for-n-in-linear-regression}

For inference to be valid in linear regression we thus need either the
error term \(\epsilon\) to be normally distributed, or the sample size
to be large enough for the central limit theorem to apply. However, what
constitutes \emph{large enough} is less clear. While the \(n\geq30\)
rule of thumb is a commonly cited lower bound for the number of
observations in regression, or parametric statistics more generally,
(see for instance Hogg, Tanis, and Zimmerman 2013; Carsey and Harden
2013; Hesterberg 2008; Cohen 1992; Wooldridge 2012; Thompson 2023) the
origin of this rule of thumb seems, at best, to be loosely related to
the distributional assumption of regression analysis. Rather, some cite
the fact that the \emph{t} distribution is nearly indistinguishable from
the normal distribution when \(\nu\geq 30\), which means that the
critical values from the normal distributions can be used rather than
having to look through a t-table (e.g Kwak and Kim 2017).\footnote{A
  highly interesting discussion on
  \href{https://stats.stackexchange.com/questions/2541/what-references-should-be-cited-to-support-using-30-as-a-large-enough-sample-siz}{Cross-Validated}
  also cites the fact that a t-table with 30 degrees of freedom neatly
  fit on a single page as a potential origin for the rule of thumb} A
keen reader will, however, note that the convergence characteristics of
the t-distribution to the normal distribution do not say anything about
whether or not the statistics of interest themselves have converged to
the \emph{t} distribution.

\(n \geq 30\) is, of course, not a universal rule of thumb; others exist
as well ranging from \(n\geq 200\) to \(n\) as a function of the number
of parameters in the model such as minimum \(n \geq 20\cdot k\) or
\(n \geq 50 + k\). These rules of thumb are often cited to ensure that
effects of certain sizes are captured, i.e.~to ensure that the
regression model has sufficient \emph{power} (for a longer discussion on
these rules of thumb, see Green 1991). However, once again, the
\emph{power} of the model does not tell us anything about the
convergence characteristics of the statistics of interest.

The \(n \geq 30\) rule is not without its critics and it has long been
noted that distributional characteristics such as the \emph{skew} affect
the utility of this rule of thumb. Already back in 2008, Hesterberg,
argued that the ``\(n \geq 30\) and not too skewed'' rule of thumb
should be retired in favor of bootstrapping techniques which are not
reliant on these distributional characteristics. Yet, Hesterberg shied
away from actually subjecting the \(n \geq 30\) rule of thumb to a
rigorous test.

Motivated by the lack of systematic evidence for the \(n \geq 30\) rule
of thumb for parametric statistics in general, Thompson (2023) embarked
on a simulation study to investigate the utility of this rule of thumb
when estimating the parameters of a set of common distributions. In this
simulation study, Thompson utilizes a set of distributional tests and a
convolutional neural network to ascertain for what \(n\) the
distribution of the parameters becomes approximately normal. The results
show that the skew of the distribution is important for that \(n\)
should be considered the lower bound.

This research note builds on a similar type of simulation design, but
focuses exclusively on regression analysis and how the distributional
characteristics of the dependent and independent variables interact with
\(n\) to affect what a reasonable lower bound of \(n\) is for regression
analysis.

\section{Simulation setup}\label{simulation-setup}

To test the distributional rules of thumb and provide additional
guidance on the lower limit of \(n\) for regression analysis, I set up a
simple simulation study. In the simulation study, a dependent variable
\(y\) is first generated from a specific distribution. An independent
variable \(x\) is then generated from its own distribution. The variable
\(y\) is generated such that the theoretical mean is zero\footnote{This
  is done by subtracting the expected value of the distribution that
  \(y\) is generated from} and independent of \(x\). This means that in
the simulation, the null hypothesis that \(\pmb{\beta} = 0\) is true.
After this, a regression model are estimated and the t-values,
i.e.~\(\hat{\beta}/SE(\hat{\beta})\), are extracted. The simulation
procedure is repeated 1,000 times to produce a distribution of estimated
t-values for each combination of \(n\) and distributions of \(y\) and
\(x\).

Since the data are generated under the null hypothesis, the t-values
extracted from the simulation should follow a t-distribution with
\(n-2\) degrees of freedom for our inferences to be valid. In other
words, if the extracted t-values follow a t-distribution, we can be
satisfied that we have a large-enough \(n\) for the distributional
assumptions of linear regression to be fulfilled.

To test if the resulting distribution of t-values actually follows a
t-distribution, I use an Anderson-Darling distributional test on the
resulting distribution and reject the null hypothesis that the
distribution belongs to the specified t-distribution if the resulting
p-value is lower than 0.05. I then repeat this procedure 500 times and
calculate the proportion of simulations in which the Anderson-Darling
test rejects the null-hypothesis that the distribution is actually a
t-distribution. As robustness tests I also estimate the less restrictive
Cramer von Mises and Kolmogorov-Smirnoff distributional tests.

To assess the severity of the violations, I also calculate the type I
error rate at \(\alpha = 0.05\) for each of the coefficients in the
regression models across each simulation. Finally, I also assess the
effects of adding a second independent variable, \(z\), to the
regression model on the rate of convergence to the t-distribution for
the t-values for \(\beta_x\) by repeating the simulation with \(z\) from
different distributions.

In this note, I am particularly interested in how distributional
characteristics such as skewness and kurtosis of \emph{both} the
dependent and independent variable(s) affect the rate of convergence to
the t-distribution for the resulting t-values. Therefore, I choose to
simulate \(y\) and \(x\) under a range of different lepto-, meso- and
platykurtic symmetric distributions as well as a range of variously
skewed distributions. I also try a large range of \(n\) ranging from 4
to 10,000.\footnote{In total I test 54 unique values of \(n\). These are
  shown in Table 2 the appendix} The tested distributions with their
corresponding skewness and kurtoses can be found in Table 1 in the
appendix. In total 6,534 different combinations of \(n\), \(y\)
distributions and \(x\) distributions are tested. Each simulation uses
500 repeats of 1,000 regression models, for a total of approximately 3.3
billion simulated regression models. In the multivariate case including
\(z\) I test an additional 37,800 combinations resulting in a total of
approximately 22.2 billion simulated regression models in the bi- and
multivariate case.\footnote{The computations were conducted on the
  Dardel PDC using resources provided by the National Academic
  Infrastructure for Supercomputing in Sweden (NAISS), partially funded
  by the Swedish Research Council through grant agreement no. 2022-06725}

\section{Results}\label{results}

The results for a select few combinations of \(x\) and \(y\)
distributions are shown in Figure 1 below.\footnote{Results for all
  combinations of distributions are available in the appendix, as are
  the results using the alternative distributional tests} These results
highlight some interesting patterns. First, it is not only when the
residuals of the regression model are normal that regression modelling
seems to perform well with a low number of observations \(n\). Rather,
as long as both \(x\) and \(y\) are \emph{symmetric} the distribution of
the t-values converges to the t-distribution already at \(n = 4\),
except in the most extreme platykurtic case (Beta 0.1,0.1) where
convergence is reached at \(n=8\). Similarly, if \emph{either} \(x\) or
\(y\) are symmetric and the other is skewed the distribution of t-values
also converge to the t-distribution at \(n = 4\) unless the skewness of
the other variable is severe (e.g.~log-normal with \(\sigma \geq 1\)).
This highlights that, contrary to theory, both the distribution of \(x\)
and \(y\) matter for the convergence of the t-values to the
t-distribution.

\begin{figure}[H]

{\centering \includegraphics{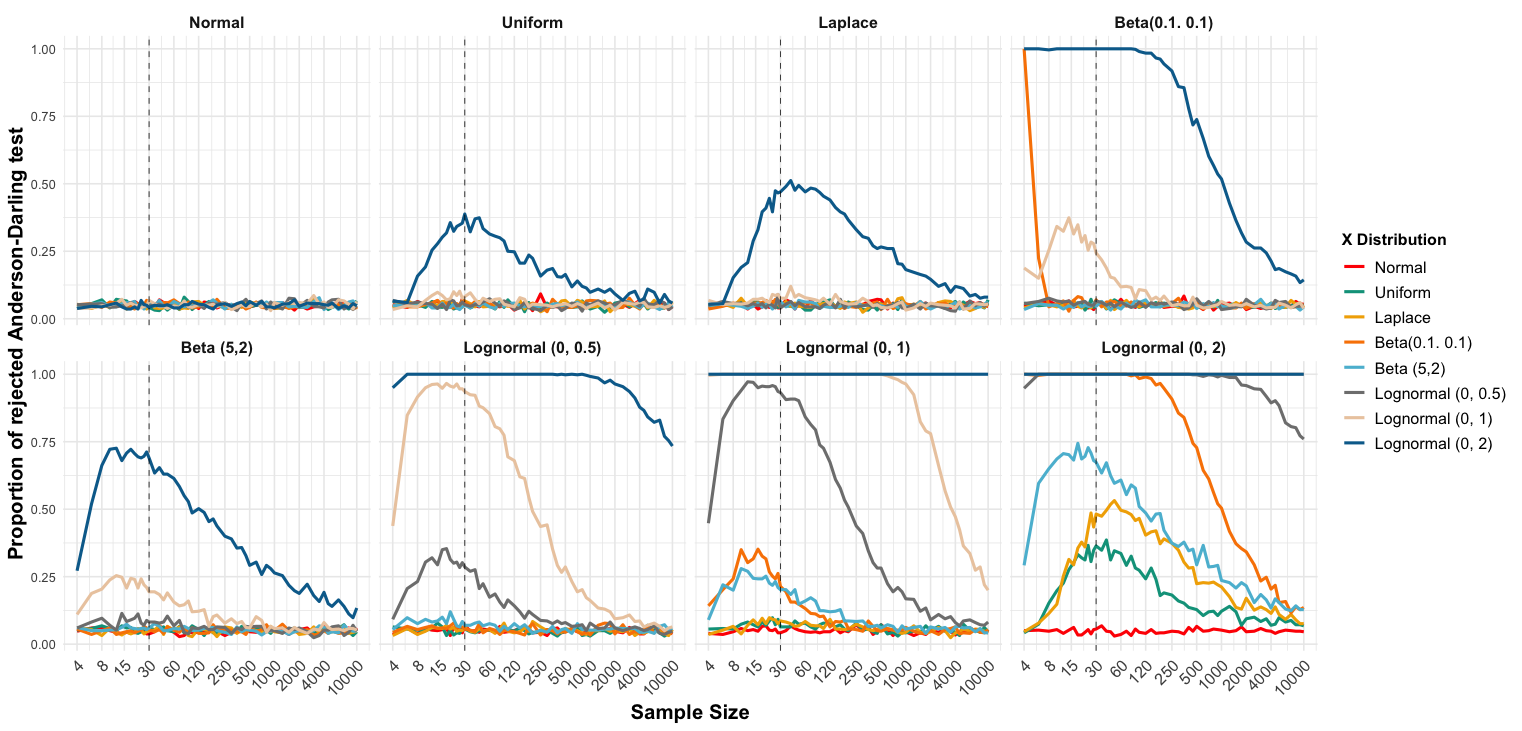}

}

\caption{Anderson-Darling rejection rate at \(\alpha = 0.05\) for
different combinations of \(x\) and \(y\) distributions and \(n\).
Facets represent different distributions of \(y\).}

\end{figure}%

When both \(y\) and \(x\) are skewed, the rate of convergence seems to
depend on both the skew of \(x\) and of \(y\). In the most extreme
cases, where both \(x\) and \(y\) follow log-normal distributions with
parameter \(\sigma \geq 1\), the t-values do not converge to the
t-distribution even at \(n = 10,000\). Except when \(y\) or \(x\) are
normal, the distribution of t-values do not converge to the
t-distribution at \(n = 10,000\) when the other variable follows a
log-normal distribution with \(\sigma = 2\).

\subsection{Effects of non-convergence to the
t-distribution}\label{effects-of-non-convergence-to-the-t-distribution}

To assess the effects of the t-values not converging to the
t-distribution I computed the type-I error rate across simulations. The
results for a select few combinations of \(x\) and \(y\) distributions
are shown in figure 2 below.\footnote{Results for all combinations of
  distributions are available in the online appendix.}. As discussed
earlier, the main concern about the t-values not converging to the
t-distribution is that the p-values obtained from the regression
analysis are no longer valid. This is especially problematic if the risk
of type-I errors is \emph{higher} than \(\alpha\) since we then risk
rejecting null hypotheses when our evidence does not support this
rejection.

\begin{figure}[H]

{\centering \includegraphics{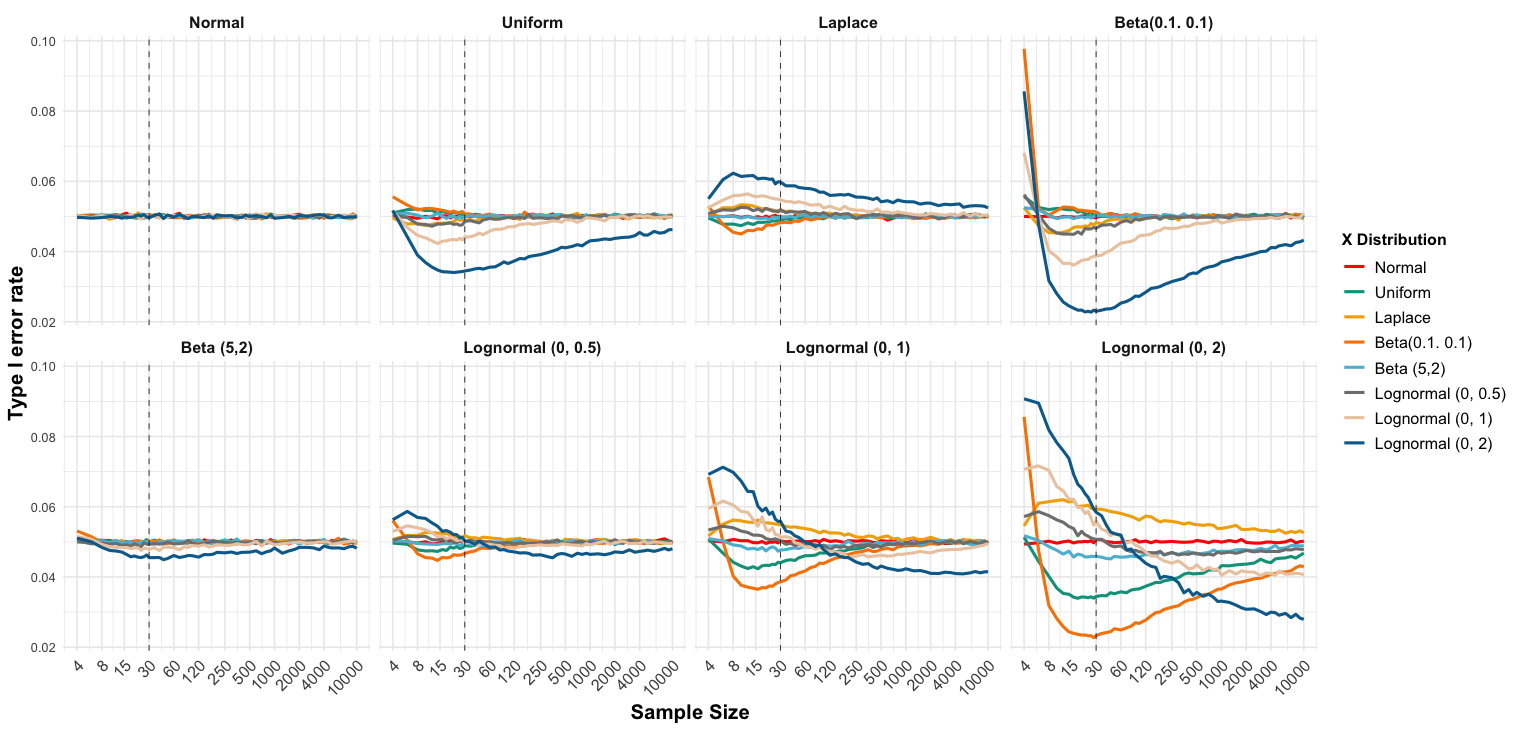}

}

\caption{Type I errors for \(\beta_1\) at \(\alpha = 0.05\) for
different combinations of \(x\) and \(y\) distributions and \(n\).
Facets represent different distributions of \(y\).}

\end{figure}%

The results from the simulation show that the consequences of the
t-values not converging to the t-distribution depends on the skewness
and kurtosis of the variables involved, and particularly the kurtosis.
For the cases where the type-I error rate is not at the appropriate
level and either of the variables are platykurtic, the type-I error rate
is \emph{lower} than \(\alpha\), except when \(n<8\). If neither of the
variables are platykurtic but are \emph{symmetric}, the type-I error
rate is \emph{higher} than \(\alpha\). The most extreme cases, where
both \(x\) and \(y\) are log-normal with \(\sigma \geq 1\), exhibit a
strange pattern where the type-I error rate is higher than alpha for low
\(n\), but lower than \(\alpha\) for higher \(n\). The results also show
that for some combinations of \(x\) and \(y\) distributions where the
Anderson-Darling test has an appropriate rejection rate, the type-I
error rate has not converged to \(\alpha\) (e.g.~when \(x\) or \(y\) are
normal and the other distribution is log-normal with \(\sigma = 1\)).

In general, however, the results also show that the effect of the
t-values not converging to the t-distribution is not as severe as one
might expect. For example, even in the worst case where both \(x\) and
\(y\) are log-normal with \(\sigma = 2\), the type-I error rate is only
approximately 0.06 when \(n =30\), i.e.~only 0.01 higher than
\(\alpha\). While the effect is larger on the platykurtic distributions,
the type-I error rate for these are \emph{lower} than \(\alpha\), which
is less problematic than if the type-I error rates are higher than
\(\alpha\) since this simply constitutes a more conservative test.

\subsection{The multivariate case}\label{the-multivariate-case}

To assess the effect of having a second independent variable in the
regression model I also ran simulations where \(z\) was added to the
regression model. \(z\) was generated under different distributions to
assess whether the skewness and kurtosis of \(z\) affected the
convergence of the t-values associated with \(\hat{\beta}_x\) to the
t-distribution. The results of this evaluation, presented in the
appendix, shows that the distribution of \(z\) did not have any
substantial effect on the convergence of the t-values associated with
\(\hat{\beta}_x\) regardless of the distribution of \(z\). This suggests
that the distribution of additional independent variables, under the
null hypothesis, does not affect the convergence of the t-values
associated with \(\hat{\beta}_x\) to the t-distribution and thus that
the results from the simulation in the bivariate case can be generalized
to the multivariate case.

\section{Discussion}\label{discussion}

This research note suggests that it is time to update what we teach our
students with regards to the lower limit of the number of observations.
Specifically, the assumption that the residuals follow a normal
distribution \emph{or} that the number of observations is sufficiently
large can actually be relaxed to that the distribution of either of the
variables involved is \textbf{symmetric} \emph{or} that the number of
observations is sufficiently large. This research note also shows that
what is \emph{sufficiently} large depends heavily on the distributional
characteristics, especially the skew, of the variables involved.

Generally, the effect of the t-values not converging to the
t-distribution is not as severe as one might expect, with the type-I
error rate typically only being slightly higher, or lower, than
\(\alpha\) in most cases. Overall, the results suggest that unless the
variables are highly skewed, the \(n\geq30\) rule of thumb seems to hold
for most combinations of distributions and may even be considered
conservative for some combinations of distributions.

What does this mean for researchers? The results suggest that it is
possible to run regressions with fewer observations than previously
thought, as long as the distribution of the variables involved are not
too skewed. This means that in situations where researchers may have
preferred to run less powerful non-parametric tests such as the
Mann-Whitney U test due to few observations and non-normality, it may be
possible to run a regression analysis instead if the variables are
symmetric, or at least not too skewed even when the number of
observations are low. At the other end of the spectrum, the results also
suggest that the \(n\geq30\) rule of thumb may be too permissive in some
cases, such as when the variables are highly skewed.

The results also suggest that the kurtosis of the variables involved
also affect what the effects of violations of the normality assumption
are. In particular, if either of the distributions of the variables are
platykurtic, the type-I error rate is lower than \(\alpha\) even when
the t-values do not converge to the t-distribution. This is less
problematic than if the type-I error rate is higher than \(\alpha\)
since this simply constitutes a more conservative test.

A general recommendation from this paper is that researchers should
report the skewness and kurtosis of the variables involved in the
regression analysis, and that the researchers should reflect on the
distributional characteristics of the variables when deciding on the
number of observations needed. If either of the distributions of the
variables are symmetric or platykurtic, fewer than 30 observations seem
in general to be sufficient. Apart from these cases, the \(n\geq30\)
rule of thumb seem to hold for most combinations of distributions except
when the distributions of either or both of the variables are highly
skewed.

That it is possible to run regressions with fewer observations than
previously thought does not mean that it is a \emph{good idea} to do so.
Power should still always be considered and the results from this
research note should not be taken as a recommendation to run regressions
with few observations. The results do, however, suggest that the
distributional assumptions needed for us to be able to trust the
t-values are not as strict as previously thought.

Further research should be directed at investigating the relationship
between skewness and convergence to the t-distribution in order to
develop guidlines for the lower \(n\) bound of regression analysis, and
to repeat this simulation study in a situation with more than one
independent variable.

\newpage{}

\section*{References}\label{references}
\addcontentsline{toc}{section}{References}

\phantomsection\label{refs}

\newpage{}

\section*{Appendix A: Included distributions and
n}\label{appendix-a-included-distributions-and-n}
\addcontentsline{toc}{section}{Appendix A: Included distributions and n}

\subsection*{Included distributions}\label{included-distributions}
\addcontentsline{toc}{subsection}{Included distributions}

\begin{table}[ht]
\centering
\begin{tabular}{lrr}
  \hline
Distribution & Theoretical skew & Theoretical kurtosis \\
  \hline
Normal & 0.00 & 3.00 \\
  Uniform & 0.00 & 1.80 \\
  Laplace & 0.00 & 6.00 \\
  Beta(0.1. 0.1) & 0.00 & 1.12 \\
  Beta (5,2) & 0.60 & 2.88 \\
  Beta (5,1) & 1.18 & 4.20 \\
  Beta (5,0.5) & 1.93 & 7.25 \\
  Lognormal (0, 0.5) & 1.75 & 8.90 \\
  Lognormal (0, 1) & 6.18 & 113.94 \\
  Lognormal (0,1.5) & 33.47 & 10078.25 \\
  Lognormal (0, 2) & 414.36 & 9220559.98 \\
   \hline
\end{tabular}
\caption{Included distributions in the simulation}
\end{table}

\subsection*{Included levels of n}\label{included-levels-of-n}
\addcontentsline{toc}{subsection}{Included levels of n}

\begin{table}[ht]
\centering
\begin{tabular}{rrrr}
  \hline
From & To & By & Number of n levels \\
  \hline
4 & 28 & 2 & 13 \\
  30 & 50 & 5 & 5 \\
  60 & 100 & 10 & 5 \\
  120 & 200 & 20 & 5 \\
  250 & 500 & 50 & 6 \\
  600 & 1000 & 100 & 5 \\
  1250 & 2000 & 250 & 4 \\
  2500 & 5000 & 500 & 6 \\
  6000 & 10000 & 1000 & 5 \\
   \hline
\end{tabular}
\caption{Included levels of n in the simulation}
\end{table}

\newpage{}

\section*{Appendix B: Full results}\label{appendix-b-full-results}
\addcontentsline{toc}{section}{Appendix B: Full results}

\subsection*{Convergence to
t-distribution}\label{convergence-to-t-distribution}
\addcontentsline{toc}{subsection}{Convergence to t-distribution}

\begin{figure}[H]

{\centering \includegraphics{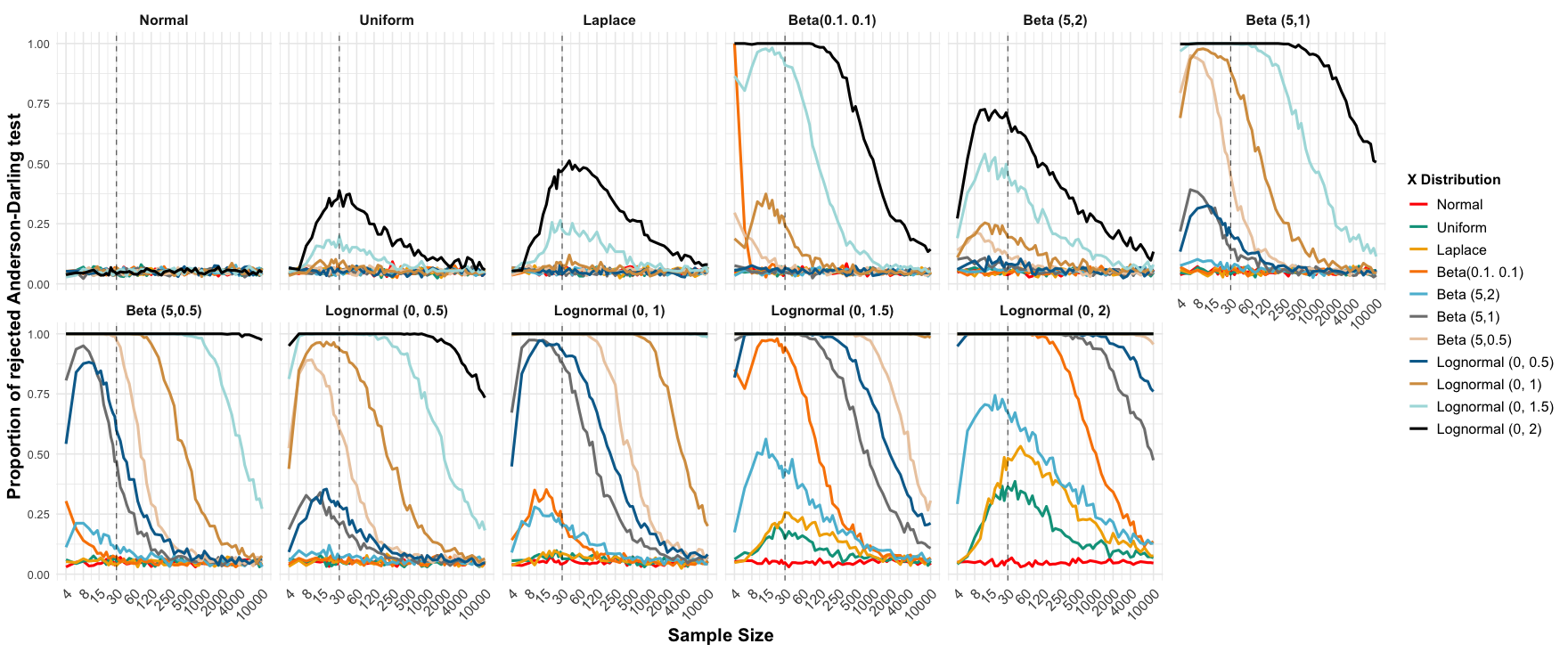}

}

\caption{Anderson-Darling rejection rate at \(\alpha = 0.05\) for
different combinations of \(x\) and \(y\) distributions and \(n\).
Facets represent different distributions of \(y\) while lines represent
different distributions of \(x\)}

\end{figure}%

\subsection*{Type I errors}\label{type-i-errors}
\addcontentsline{toc}{subsection}{Type I errors}

\begin{figure}[H]

{\centering \includegraphics{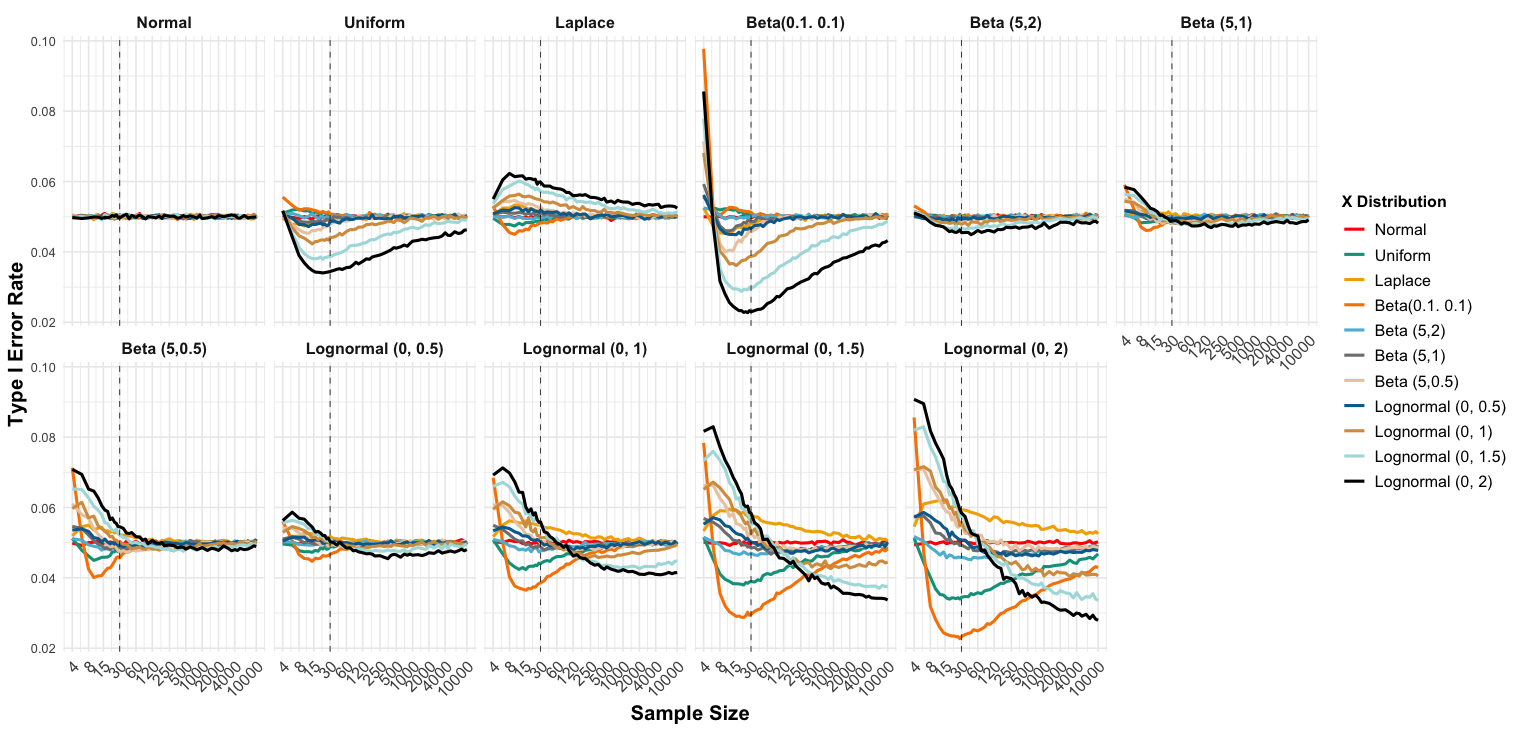}

}

\caption{Type I errors for \(\beta_1\) at \(\alpha = 0.05\) for
different combinations of \(x\) and \(y\) distributions and \(n\).
Facets represent different distributions of \(y\) while lines represent
different distributions of \(x\)}

\end{figure}%

\subsection*{Alternative distributional
tests}\label{alternative-distributional-tests}
\addcontentsline{toc}{subsection}{Alternative distributional tests}

\subsubsection*{Cramer-von Mises test}\label{cramer-von-mises-test}
\addcontentsline{toc}{subsubsection}{Cramer-von Mises test}

\begin{figure}[H]

{\centering \includegraphics{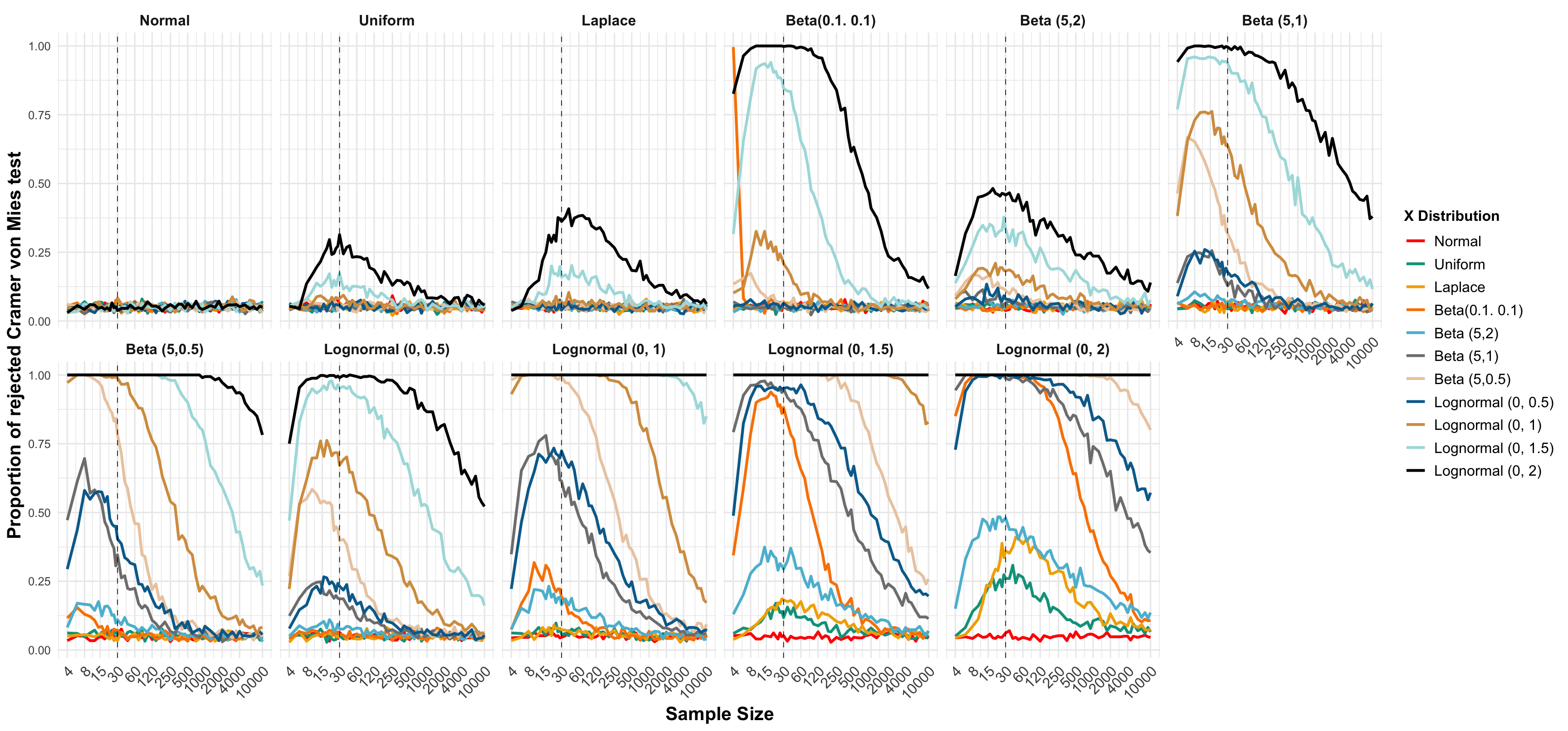}

}

\caption{Cramer-von Mises rejection rate at \(\alpha = 0.05\) for
different combinations of \(x\) and \(y\) distributions and \(n\).
Facets represent different distributions of \(y\) while lines represent
different distributions of \(x\)}

\end{figure}%

\subsubsection*{Kolmogorov-Smirnoff
test}\label{kolmogorov-smirnoff-test}
\addcontentsline{toc}{subsubsection}{Kolmogorov-Smirnoff test}

\begin{figure}[H]

{\centering \includegraphics{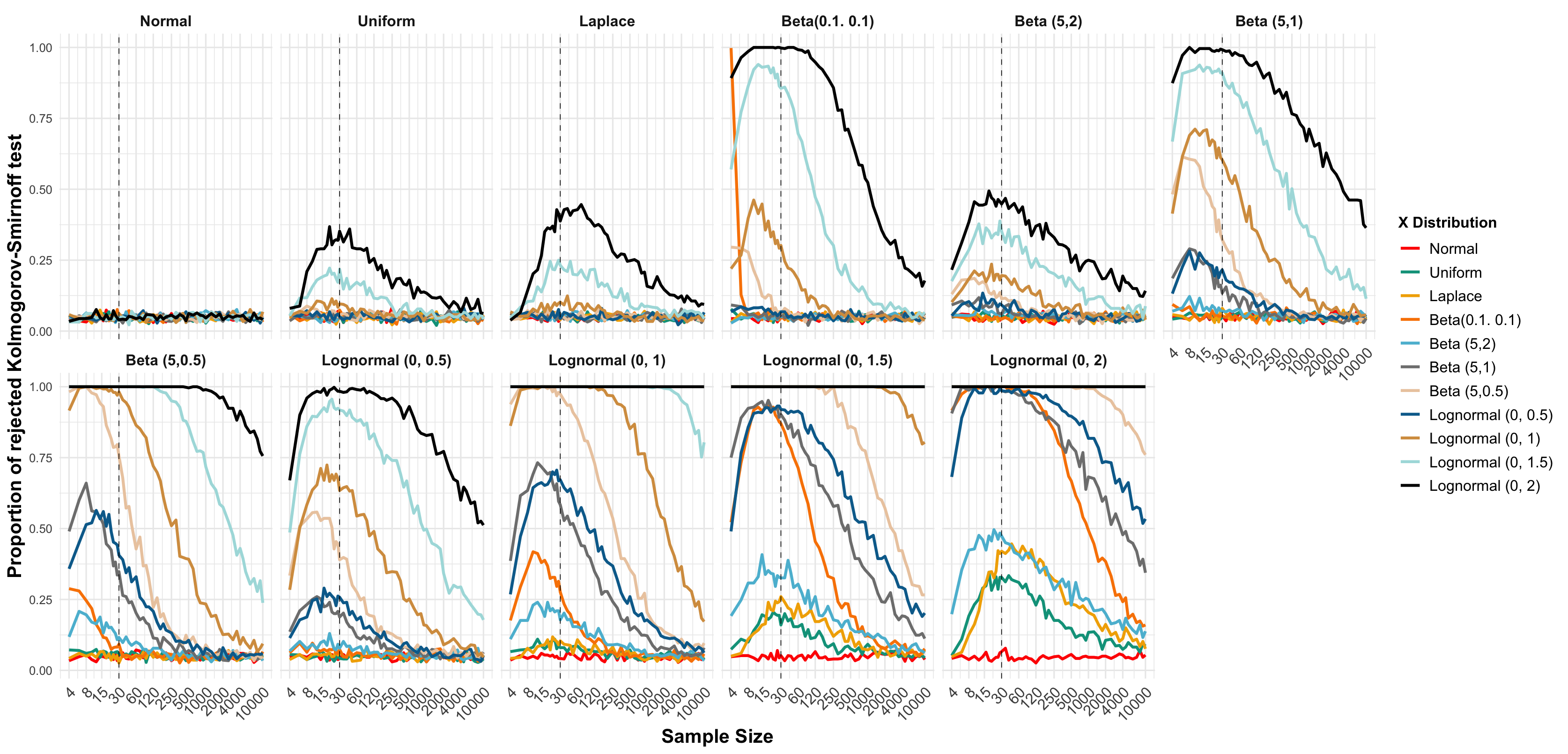}

}

\caption{Kolmogorov-Smirnoff rejection rate at \(\alpha = 0.05\) for
different combinations of \(x\) and \(y\) distributions and \(n\).
Facets represent different distributions of \(y\) while lines represent
different distributions of \(x\)}

\end{figure}%

\subsection*{The multivariate case}\label{the-multivariate-case-1}
\addcontentsline{toc}{subsection}{The multivariate case}

\begin{figure}[H]

{\centering \includegraphics[width=\textwidth,height=0.8\textheight]{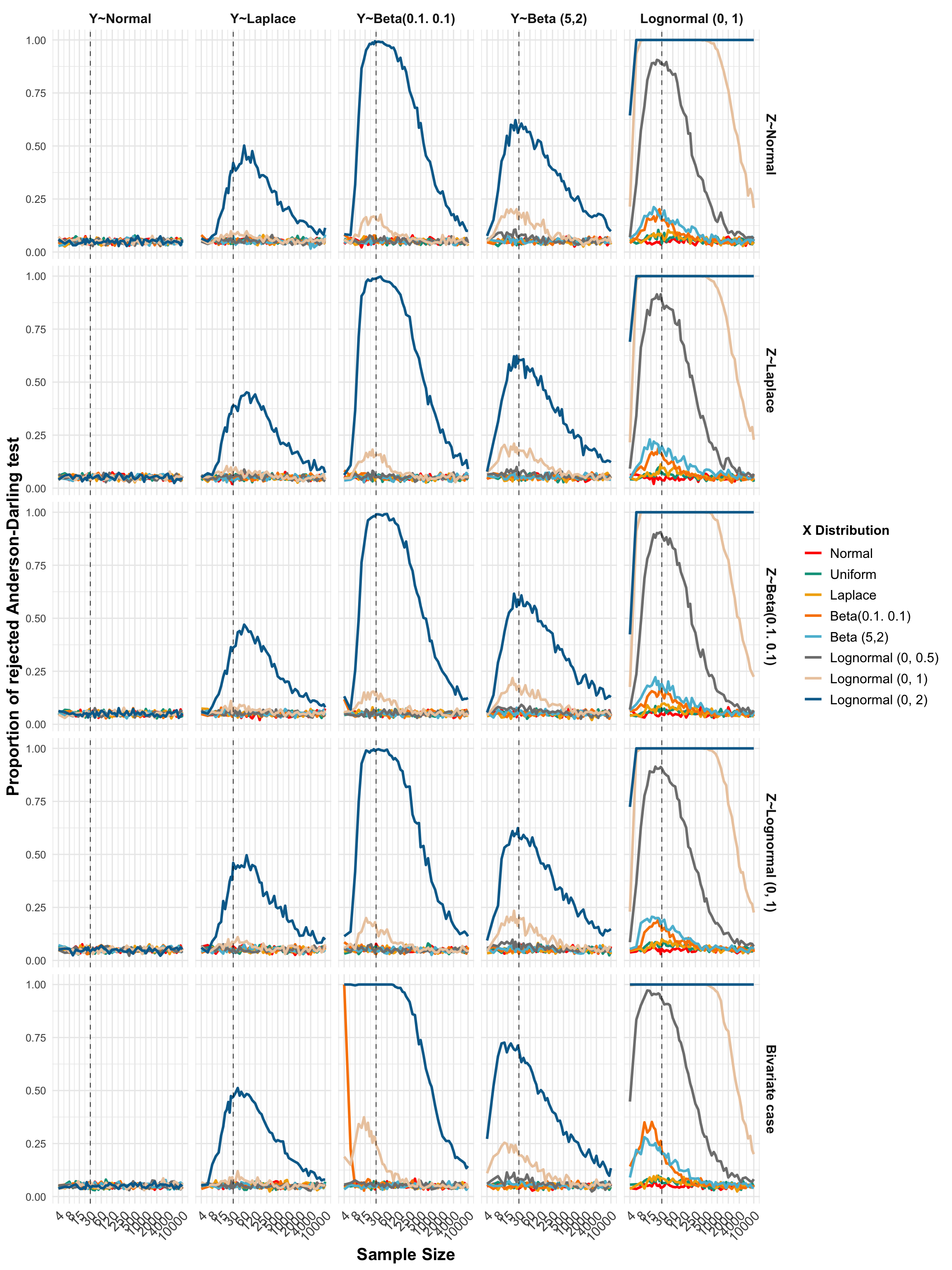}

}

\caption{Anderson-Darling rejection rate at \(\alpha = 0.05\) for
different combinations of \(x\), \(y\), and \(z\) distributions and
\(n\). Columns represent different distributions of \(y\), rows
different distributions of \(z\) and lines represent different
distributions of \(x\). Only a subset of combinations of distributions
of \(y\) and \(z\) are included in the figure. Results for the remaining
combinations of distributions, and the Kolmogorov-Smirnoff and Cramer
von Mies rejection rates, can be made available on request.}

\end{figure}
\begin{figure}[H]

{\centering \includegraphics[width=\textwidth,height=0.9\textheight]{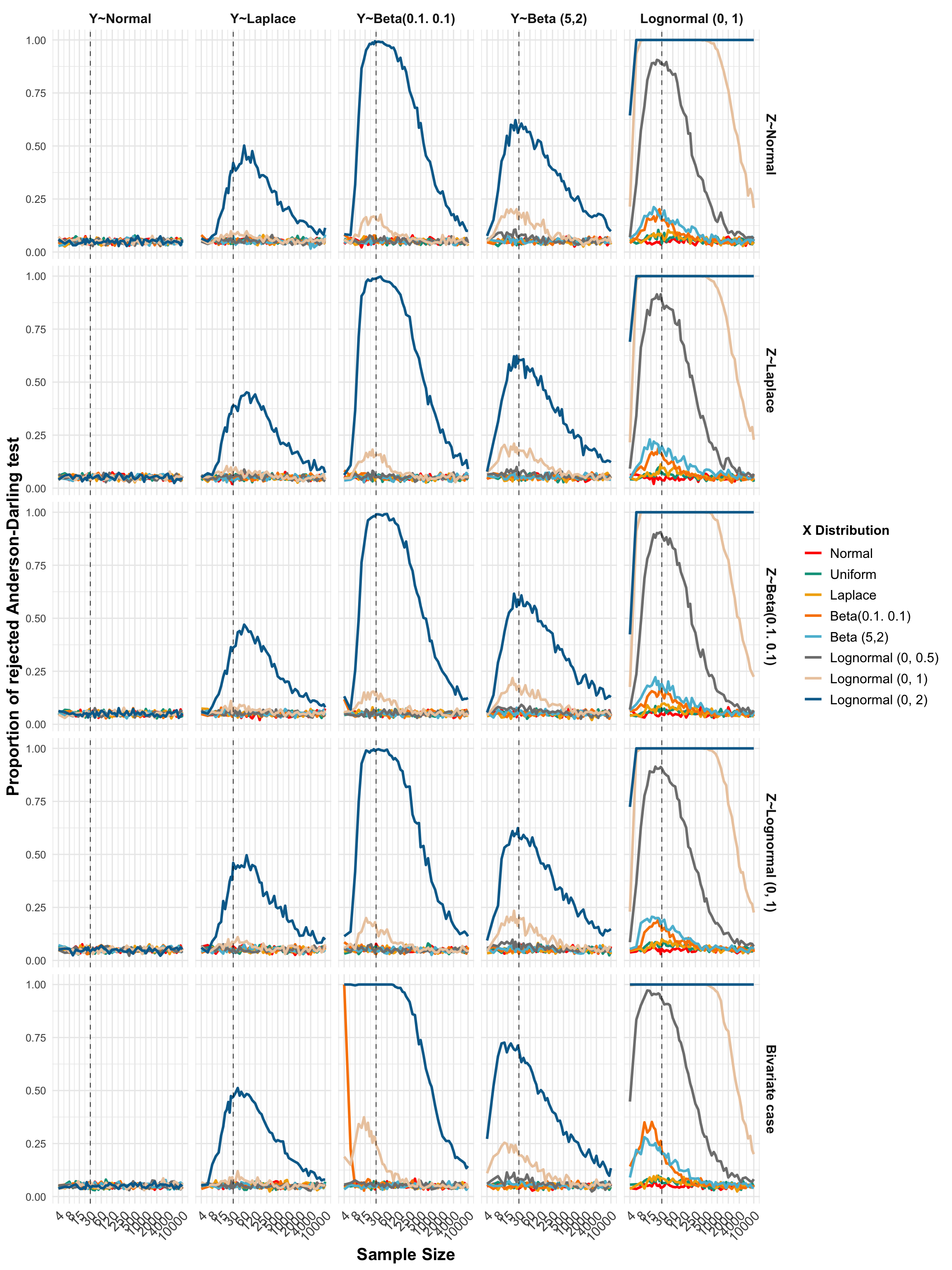}

}

\caption{Type I Error rates at \(\alpha = 0.05\) for different
combinations of \(x\), \(y\), and \(z\) distributions and \(n\). Columns
represent different distributions of \(y\), rows different distributions
of \(z\) and lines represent different distributions of \(x\). Only a
subset of combinations of distributions of \(y\) and \(z\) are included
in the figure. Results for the remaining combinations of distributions
can be made available on request.}

\end{figure}%

\end{document}